\documentclass[epj,referee]{svjour}
\usepackage{latexsym}
\usepackage{graphics}
\usepackage{array}
\usepackage{amssymb,amsfonts,amsmath}
\usepackage{bm}
\usepackage[breaklinks]{hyperref}
\usepackage[all]{hypcap}

\newcommand{\bb}[0]{\begin{eqnarray}}
\newcommand{\ee}[0]{\end{eqnarray}}
\newcommand{\eref}[1]{Eq.~(\ref{#1})}
\newcommand{\efig}[1]{Fig.~\ref{#1}}
\newcommand{\pb}{p_0}
\newcommand{\qb}{q_0}
\begin{document}
\title{De Haas-van Alphen oscillations in the compensated organic metal $\alpha$-'pseudo-$\kappa$'-(ET)$_4$H$_3$O[Fe(C$_2$O$_4$)$_3$]$\cdot$(C$_6$H$_4$Br$_2$)}
\author{Alain Audouard\inst{1}\thanks{\emph{e-mail: alain.audouard@lncmi.cnrs.fr}} \and Jean-Yves~Fortin\inst{2}\thanks{\emph{e-mail: fortin@ijl.nancy-universite.fr}} \and Vladimir N. Laukhin\inst{3,4} \and David Vignolles\inst{1} \and
Tatyana G. Prokhorova\inst{5} \and Eduard B. Yagubskii\inst{5} \and Enric Canadell\inst{4}
}                     
%
%
\institute{Laboratoire National des Champs Magn\'{e}tiques
Intenses (UPR 3228 CNRS, INSA, UJF, UPS) 143 avenue de Rangueil,
F-31400 Toulouse, France. \and Institut Jean Lamour, D\'epartement de Physique de la
Mati\`ere et des Mat\'eriaux,
CNRS-UMR 7198, Vandoeuvre-les-Nancy, F-54506, France. \and Instituci\'{o} Catalana de Recerca i Estudis Avan\c{c}ats (ICREA), 08010 Barcelona, Spain. \and Institut de Ci\`{e}ncia de Materials de Barcelona, CSIC, Campus de
la UAB, 08193, Bellaterra, Spain. \and Institute of Problems of Chemical Physics, Russian Academy of Sciences, 142432 Chernogolovka, MD, Russia}
\date{Received: date / Revised version: date}
%
\abstract{
Field-, temperature- and angle-dependent Fourier amplitude of de Haas-van Alphen (dHvA) oscillations are calculated for compensated two-dimensional (2D) metals with
textbook Fermi surface (FS) composed of one hole and two electron orbits connected by magnetic breakdown. It is demonstrated that, taking into account the opposite sign
of electron and hole orbits, a given Fourier component involves combination of several orbits, the contribution of which must be included in the calculations. Such FS
is observed in the strongly 2D organic metal $\alpha$-'pseudo-$\kappa$'-(ET)$_4$H$_3$O[Fe(C$_2$O$_4$)$_3$]$\cdot$(C$_6$H$_4$Br$_2$), dHvA oscillations of which have
been studied up to 55 T for various directions of the magnetic field with respect to the conducting plane. Calculations are in good quantitative agreement with the data.
} 

\authorrunning{Alain Audouard et al.}
\titlerunning{De Haas-van Alphen oscillations in a compensated metal}

\maketitle
\section{Introduction}
\label{intro}
Provided that no phase transition occurs as the temperature is lowered, Fermi surface (FS) of two-dimensional (2D) organic metals is generally rather simple and achieve model systems for quantum oscillations physics. Indeed, in the numerous cases where the compound possesses two carriers (generally holes) per unit cell, the FS originates from a single orbit with an area equal to that of the first Brillouin zone \cite{Ro04}. In an extended zone scheme, these orbits overlap either along (i) one or (ii) two directions yielding in magnetic field either (i) the model linear chain of orbits coupled by magnetic breakdown (MB) proposed by Pippard in the early sixties \cite{Pi62,Sh84} for which all the orbits are of the same (hole) type or (ii) a set of compensated electron- and hole-type orbits (i.e. the sum of the hole-type orbits cross section is equal to that of the electron-type orbits).

The former case (i) has been  widely studied: it is known that quantum oscillations spectra are strongly affected by field-induced chemical potential oscillations yielding many frequency combinations, the field and temperature dependence of which cannot be accounted for by the Lifshitz-Kosevich (LK) formula \cite{Wo96,Si00,Au13}. In contrast, to our best knowledge, the second case (ii) has only received little attention up to now even though MB between hole- and electron-type orbits is relevant, for instance, for recently studied oxide superconductors \cite{Ca10,He10,He14}. Numerical resolution of the grand potential equation for the case (ii) reveals that
chemical potential oscillations are strongly damped for compensated orbits, even in the case of 2D metals \cite{Fo08,Fo09}. As a consequence, contrary to case (i), the
LK formula is predicted to account for the field and temperature dependence of the oscillations amplitude in case (ii). However, depending on the MB probability and taking into account the opposite sign of electron and hole orbits, a given Fourier component can involve combinations of an infinite set of orbits, the contribution of which must be included in the Fourier amplitudes calculation. From the experimental side, only few quasi-2D compensated metals have been synthesized yet. Such FS are observed in e.g. (BEDO-TTF)$_2$ReO$_4\cdot$H$_2$O \cite{Kh98} (where BEDO-TTF stands for the bis-ethylenedioxy-tetrathiafulvalene molecule) and, more recently synthesized, (ET)$_4$H$_3$O[Fe(C$_2$O$_4$)$_3$]$\cdot$ Solv (where ET stands for the bis-ethylenedithio-tetrathiafulvalene molecule and Solv is an organic solvent)  \cite{Zo11,Zo12}.

\begin{figure}
\centering
\resizebox{0.9\columnwidth}{!}{
  \includegraphics{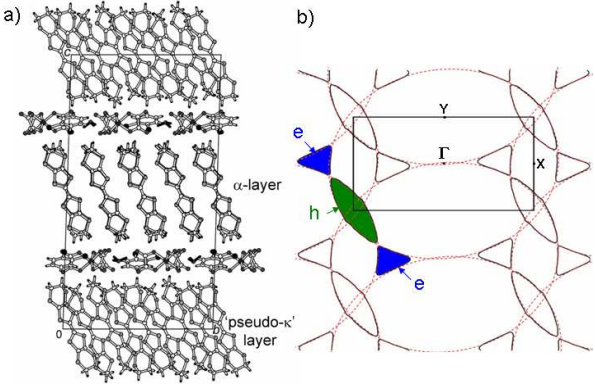}
}
\caption{\label{Fig:structures} (color on line) (a) Crystal structure of $\alpha$-'pseudo-$\kappa$'-(ET)$_4$H$_3$O[Fe(C$_2$O$_4$)$_3$]$\cdot$(C$_6$H$_4$Br$_2$). $\alpha$-type and 'pseudo-$\kappa$'-type layers are conducting and insulating, respectively. (b) Fermi surface relevant to the $\alpha$-type layers, according to Zorina et al.\cite{Zo11}. Blue and green areas mark electron- and hole-type orbits, respectively. Ellipses in dashed lines mark the orbits from which the Fermi surface originates. Labels $\Gamma$, X and Y refer to  the (0, 0), ($a^*$/2, 0) and (0, $b^*$/2)  points of the first Brillouin zone, displayed as a rectangle in solid line.  }
\end{figure}

As reported in Fig.~\ref{Fig:structures}(a), the unit cell of $\alpha$-'pseudo-$\kappa$'-(ET)$_4$H$_3$O[Fe(C$_2$O$_4$)$_3$]$\cdot$(C$_6$H$_4$Br$_2$) \cite{Zo11} contains two different donor planes with different packing. One of them, with a 'pseudo-$\kappa$' structure is insulating while the other with an $\alpha$-type structure is metallic. As a consequence, spacing between conducting layers is as large as 3.64 nm, ensuring negligibly small interlayer transfer integral, hence avoiding effects due to FS corrugation \cite{Gr02}. According to band structure calculations, the FS of the metallic plane, displayed in Fig.~\ref{Fig:structures}(b), is composed of two electron-type and one hole-type compensated orbits (i.e. the hole orbit area is twice the electron orbits area). In the extended zone scheme, this set of three orbits is isolated from the other sets. In other words, we are not dealing with a network of coupled orbits. In contrast, each of the orbits within a set is liable to be connected to the other by MB. This feature is shared by the FS studied in Ref. \cite{Fo08}, which is composed of one hole and one electron compensated orbit.

The aim of this paper is, in the first step, to provide calculation of the Fourier amplitudes of de Haas-van Alphen (dHvA) oscillations spectra for the FS of Fig.~\ref{Fig:structures}(b). In the second step, magnetic torque oscillations of $\alpha$-'pseudo-$\kappa$'-(ET)$_4$H$_3$O[Fe(C$_2$O$_4$)$_3$]$\cdot$(C$_6$H$_4$Br$_2$), yielding dHvA spectra, are studied in magnetic fields up to 55 T. Whereas magnetoresistance data of this organic metal, measured in magnetic fields below 16 T reveal only one frequency attributed to the hole orbit \cite{Zo11}, the two frequencies corresponding to electron and hole orbits area are observed. It is shown that the field and temperature dependence of the Fourier amplitudes are in agreement with the reported calculations.

\section{Experimental}

Crystals were synthesized by electrocrystallization technique as reported by Zorina et al.\cite{Zo11}.  Two crystals denoted as crystal \#1 and \#2 hereafter were studied. Their approximate dimensions were 0.12 $\times$ 0.1 $\times$ 0.04~mm$^3$. Magnetic torque was
measured  with a commercial piezoresistive microcantilever, in pulsed magnetic
fields of up to 55 T with a pulse decay duration of 0.32 s. Variations of the
cantilever piezoresistance were measured in the temperature range from 1.4 K to
4.2 K with a Wheatstone bridge with an $ac$ excitation at a frequency of 63 kHz.
The angle between the normal to the conducting plane and the magnetic field
direction was $\theta$ = 15$^{\circ}$ and 32$^{\circ}$ for crystal \#1 while $\theta$ was varied from 15$^{\circ}$ to 71$^{\circ}$ thanks to a rotating sample holder for crystal \#2.

\section{Calculations of the Fourier amplitudes}
\label{Sec:calculations}
In the framework of the Lifshitz-Kosevich and Falicov-Stachowiak
models~\cite{Sh84,Fa66}, the oscillatory part of the magnetization
$M_{osc}$ for a set of 2D orbits $\eta$ can be written as

\begin{eqnarray}
\label{eq:Mosc}
M_{osc}=-\sum_{\eta}\sum_p \frac{F_{\eta}}{\pi}A_{p\eta}
\sin(2\pi p\frac{F_{\eta}}{B\cos\theta})
\end{eqnarray}

where $\theta$ is the
angle between the normal to the conducting plane and the field direction. The index $\eta$ stands for all the closed orbits allowed by the FS topology, including MB orbits, with fundamental frequencies $F_{\eta}$. They do not include orbits which can be associated
with an harmonic of a simpler trajectory, the latter being accounted for by the index $pF_{\eta}$ (with $p >$ 1)  where $p$ is
the harmonic order. The amplitude $A_{p\eta}$
of the Fourier component with frequency $pF_{\eta}$ depend
on parameters such as temperature, magnetic field, effective masses ($m_{\eta}$),
Dingle temperatures ($T_{D\eta}$), MB field ($B_0$), and effective Land\'e factors ($g^*_{\eta}$). These amplitudes can
be expressed as $A_{p\eta}=(-1)^{ps_{\eta}}R^{MB}_{p\eta}R_{p\eta}/(pm_{\eta})$. MB damping factor is given by $R^{MB}_{p\eta}=C_{p\eta}(i\pb)^{pt_{\eta}}\qb^{pb_{\eta}}$, where $C_{p\eta}$ is the symmetry factor of the orbit $p\eta$, $t_{\eta}$ and
$b_{\eta}$ are the number of tunnelings and reflections, respectively,
encountered by a quasi-particle during its path and $2s_{\eta}$ is the number
of turning points around the orbit $\eta$. The tunneling and reflection
probabilities at a MB junction are given, in agreement with the Chambers approximation,
by $\pb^2=\exp(-B_0/B\cos\theta)$ and $\qb^2=1-\pb^2$, respectively \cite{Ch66}. The damping factor
$R_{p\eta}$ can be written as the product of thermal, Dingle and spin damping
factors ($R_{p\eta}=R^T_{p\eta}R^D_{p\eta}R^s_{p\eta}$) which
are given by $R^T_{p\eta}=X_{p\eta}/\sinh(X_{p\eta})$,
$R^D_{p\eta}=\exp(-u_0T_{D\eta}pm_{\eta}/B\cos\theta)$, and
$R^s_{p\eta}=\cos(p\pi g^*_{\eta}m_{\eta}/2\cos\theta)$, respectively, where
$X_{p\eta}=u_0Tpm_{\eta}/B\cos\theta$ and $u_0=2\pi^2k_B^2m_e/e\hbar$.

\begin{figure}
\centering
\resizebox{0.75\columnwidth}{!}{
  \includegraphics{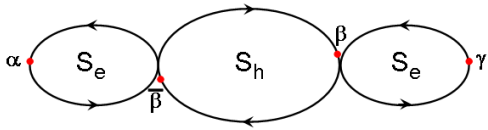}
}
\caption{\label{Fig:model} Compensated Fermi surface with classical
representation of the quasi-particle orbits. Phase variation of the
wave-function is given by $\exp(iS_e)$ around the electronic orbit $e$ and
$\exp(iS_h)$ around the hole orbit $h$, where $S_{e(h)}$ is the electron (hole) surface area. Either magnetic breakdown occurs with the
probability amplitude $i\pb$ between two orbits or reflection is allowed
with the probability amplitude $\qb=(1-\pb^2)^{1/2}$.}
\end{figure}

In the case of the FS of Fig.\ref{Fig:structures}(b), which is modeled by
\efig{Fig:model}, and owing to the orbits
compensation predicted by band structure calculations, each electron ($e$) and
hole ($h$) orbit contributes to the dHvA oscillations spectrum with the
frequency $F_e$ and $F_h$ = 2$F_e$, respectively. Besides, MB orbits composed of
several individual orbits ($\eta = n_ee+n_hh$) are also liable to contribute to
the spectrum. Indeed, all contributing trajectories for a given frequency are
accounted for by considering the amplitude and
phase variation $\exp(iS_{\eta})=\exp[2i\pi h{\cal A_{\eta}}/eB)]$ of the
wave-function, where ${\cal A_{\eta}}$ is the area of the trajectory in
the Brillouin zone, which is directly identified to the frequency $F_{\eta}=|{\cal
A_{\eta}}|h/e$. Taking into account the opposite sign of electron and hole
surface area, $S_{n_ee+n_hh}=n_eS_e+n_hS_h$ with $S_h=-2S_e$, all Fourier
components with frequencies
$F_{n_ee+n_hh}=|n_eF_e-n_hF_h|=|n_e-2n_h|F_e$, and effective masses
$m_{n_ee+n_hh}=n_hm_h+n_em_e$~\cite{Sh84,Fa66}, where $n_e$ and $n_h$ are the
(positive) numbers of electron and hole orbits involved, contribute. Therefore, the
dominant frequencies $F_e$, $2F_e=F_h$ and $3F_e$  we will consider in the following,
arise from infinite orbits combinations with $n_e =
2n_h\pm1$, $n_e = 2n_h\pm2$ and $n_e = 2n_h\pm3$, respectively.
To obtain the amplitude of these Fourier components, the Fermi surface is modeled in
\efig{Fig:model} by a linear finite chain of $e$ and $h$ orbits.

To count all the allowed paths from one of the arbitrary starting and ending
points $\alpha,\bar\beta,\beta,\gamma$ in \efig{Fig:model}, we consider paths
of length $n$, where $n$ is an integer corresponding to the number of
elementary steps from one
point to another that form and define the complete closed trajectory. Turning
points are located at the arrows in \efig{Fig:model} and going through one
of these is tantamount to adding a factor $i$ in the wave-function amplitude.
We introduce the vector amplitude $|\alpha,\bar\beta,\beta,\gamma\rangle$
corresponding to the state of the quasi-particle. Any closed path is therefore defined
by its length $n$ and its starting point and state
$s_0$ among the set ($\alpha,\bar\beta,\beta,\gamma$). Initially the state
$|s_0\rangle$ is filled with zeroes except for the
component corresponding to $s_0$, for example
$|\alpha\rangle=|1,0,0,0\rangle$. We then define each step factor of the
trajectory by moving from one point
among the set ($\alpha,\bar\beta,\beta,\gamma$) to the next one, by
following the direction imposed by the field, with the adequate changes in the
phase and amplitude. For example, as shown in \efig{Fig:model}, $\beta=-i\pb\alpha\exp(iS_e/2+iS_h/2)+i\qb\bar\beta\exp(iS_h/2)$.
Introducing parameters $x=\exp(iS_e/2)$ and
$y=\exp(iS_h/2)$, with $y=1/x^2$, we can write a transfer matrix for all the
elementary steps on the Fermi surface:

\[
T(x,y)=
\begin{pmatrix}
-\qb x^2& -i\pb xy & 0 & 0  \\
 0 & 0 & i\qb y & -\pb x \\
-\pb x & i\qb y & 0 & 0 \\
0 & 0 & -i\pb xy & -\qb x^2  \\
\end{pmatrix}
\]

Since we consider closed paths and conserved
current, the output vector is the same as the input vector $|s_0\rangle$. Then the number of all possible paths of length $n$ and starting from a point $s_0$ is given
by the number $\langle s_0|T(x,y)^{n}|s_0\rangle$. The generalized damping factor $A_{p\eta}$ in \eref{eq:Mosc} is computed for all possible orbit combinations contributing to the same frequency $pF_{\eta}$. The combinatorial factor $(-1)^{ps_{\eta}}$ $R^{MB}_{p\eta}$ in $A_{p\eta}$, combining the different ways of drawing the orbits
$p\eta$ on the FS, is precisely defined as
the coefficient of $x^{2n_e}y^{2n_h}$, as well as $x^{-2n_e}y^{-2n_h}$, in the
following polynomial function

\bb
A(x,y)=\int_0^1\frac{dz}{z}\sum_{s_0}\sum_{n\ge 0}W_{s_0}\langle
s_0|T^n(\sqrt{z}x,zy)|s_0\rangle
\ee

weighted by damping factors for each orbit, and corresponding
to integers $(n_e,n_h)$. $W_{s_0}$ is the $weight$ of the
point $s_0$ in the quasi-particle path. We will take $W_{\alpha}=W_{\gamma}=1$
and $W_{\beta}=W_{\bar\beta}=1/2$.
Indeed points $\beta$ and $\bar\beta$ are equivalent in the sense that they define
the same set of paths and belong to the same orbit. Finally, the
integration over $z$ is performed so that to remove the number of cyclic
permutations $n_e+2n_h$ of the same fundamental orbit by adding a
compensatory factor $1/(n_e+2n_h)$, and
to take into account the harmonics coefficients with a weighting factor $1/p$. Each path can be simply
decomposed using $e$ (or $x$) for a portion of trajectory around the electron orbits, and $\sqrt{h}$ (or $y$) around the hole orbit, and $z$ is added for
each $e$ or $\sqrt{h}$ encountered on the trajectory. For example the orbit
$ee\sqrt{h}e\sqrt{h}$ (corresponding to $n_e=3,\;n_h=1$) with effective mass $3m_e+m_h$ and
phase $S_e$ (corresponding to the frequency $F_e$)
has a weight proportional to $10\pb^4\qb/5=2\pb^4\qb$ (the factor $5=n_e+2n_h$
is the
number of cyclic permutations of the same
orbit $ee\sqrt{h}e\sqrt{h}=e\sqrt{h}e\sqrt{h}e=\cdots$). It is different
from the orbit $ee\sqrt{h}\sqrt{h}e$ for example, which has a weight proportional to
$-2\pb^3\qb^2$. We find that the
first three amplitudes can be written as

\bb\label{eq:Ae}
\nonumber
A_{e}&=&-\frac{2\qb}{m_e} R_e-\frac{2\pb^2\qb}{m_e+m_h} R_{e+h}
\ee
\bb
+\frac{2[\pb^4\qb-\pb^2\qb^3]}{3m_e+m_h}R_{3e+h}+\cdots,
\ee
\bb
\label{eq:A2e}
\nonumber
A_{2e}&=&\frac{\qb^2}{2m_e}R_{2e}-\frac{\qb^2}{m_h}R_{h}
\ee
\bb
+\frac{3\pb^4\qb^2-2\pb^2\qb^4}{2(m_e+m_h)}R_{2e+2h}
+\frac{2\pb^2\qb^4-3\pb^4\qb^2}{4m_e+m_h}R_{4e+h}+\cdots,
\ee
\bb
\label{eq:A3e}
A_{3e}&=&-\frac{2\qb^3}{9m_e}R_{3e}
+\frac{2\pb^2\qb^3}{m_e+2m_h}R_{e+2h}+\cdots.
\ee

where $R_{p\eta}=R^T_{p\eta}R^D_{p\eta}R^s_{p\eta}$. As an example, the first term $-(2\qb^3/9m_e)R_{3e}$ of $A_{3e}$ in Eq.~\ref{eq:A3e}, comes
from the third harmonics of the $e$ orbit. Indeed, according to Eq.~\ref{eq:Mosc}, this factor is given by $F_e(-2\qb^3)R_{3e}/(3m_e)$ which can be
rewritten as $3F_e(-2\qb^3)R_{3e}/(9m_e)$. Magnetization can then be expanded as

\bb\label{eq:Mosc2}
\nonumber
M_{osc}=-\frac{F_{e}}{\pi}A_{e}\sin\left ( 2\pi\frac{F_{e}}{B}\right )
-\frac{2F_{e}}{\pi}
A_{2e}\sin\left (2\pi\frac{2F_{e}}{B}\right )
\ee
\bb
-\frac{3F_{e}}{\pi}
A_{3e}\sin\left (2\pi\frac{3F_{e}}{B}\right )
+\cdots
\ee

\begin{figure*} 
\centering
\resizebox{1.0\columnwidth}{!}{
  \includegraphics{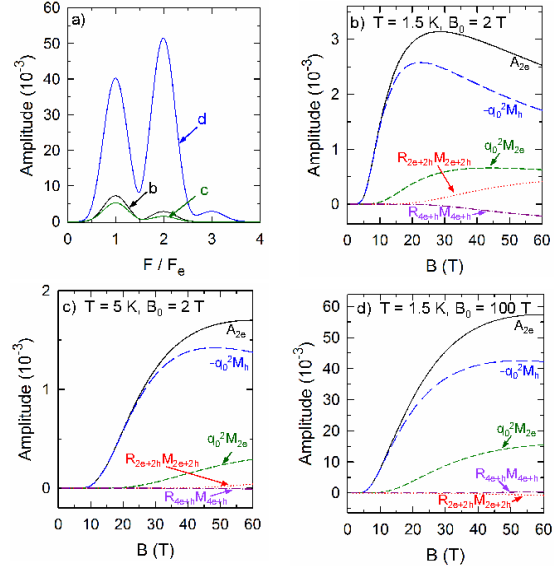}
}
\caption{\label{Fig:calculs} (color on line) (a) Fourier spectra of the oscillatory magnetization calculated at a mean field of 40 T from Eq.~\ref{eq:Mosc2} with  $m_e$ = 1, $m_h$ = 1, $g^*_e$ = 2,  $g^*_h$ = 2, $T_{De}$ = 1 K, $T_{Dh}$ = 1 K. Labels $b, c, d$ correspond to T = 1.5 K and $B_0$ = 2 T, T = 5 K and $B_0$ = 2 T, T = 1.5 K and $B_0$ = 100 T, respectively. (b), (c), (d)  Field dependence of the amplitude of the Fourier component with frequency $F_h= 2F_e$ corresponding to Labels $b, c, d$, respectively, of (a). Solid lines are calculated from Eq.~\ref{eq:A2e}. Long dashed, short dashed, dotted and dash-dotted lines are the components linked to the $h$, $2e$, $2e+2h$ and $4e+h$ orbits, respectively. The magnetic breakdown damping factors for the orbits $2e+2h$ and $4e+h$ are given by $R_{2e+2h}=3p_0^4q_0^2-2p_0^2q_0^4$ and $R_{4e+h}=2p_0^2q_0^4-3p_0^4q_0^2$, where $q_0$ and $p_0$ are the reflection and magnetic breakdown probability, respectively. }
\end{figure*}

Examples of Fourier spectra deduced from Eq.~\ref{eq:Mosc2} are reported in Fig.~\ref{Fig:calculs}(a) for various temperatures and MB fields. Land\'{e} factors ($g^*_e$ = 2,  $g^*_h$ = 2) and effective masses ($m_e$ = 1, $m_h$ = 1), which otherwise are close to those deduced from the data reported in the next section, are chosen so that the absolute value of the spin damping factors is equal to 1 ($| R^s_{p\eta}|$=1) in order to avoid any spurious effect due to spin-zero phenomenon. First, $A_{3e}$ is always small compared to $A_{e}$ and $A_{2e}$. Corresponding field-dependent amplitudes $A_{2e}$ are given in Figs.~\ref{Fig:calculs}(b), (c), (d). Despite the effective mass $m_{\eta}$ of a given $\eta$ orbit increases as the number of individual orbits involved increases, a clear contribution of the orbits $2e$, $2e+2h$ and $4e+h$ to the amplitude is observed in Fig.~\ref{Fig:calculs}(b). Their relative contributions decrease as the Dingle temperatures (not shown), the temperature (see Fig.~\ref{Fig:calculs}(c)) and the MB field (see Fig.~\ref{Fig:calculs}(d)) increase
which may lead to errors in effective mass determination \cite{Fo09}.
Contribution of the orbit $2e$ is substantial in any case, which indicates that
this orbit needs to be considered, even in the case of large MB field,
scattering rate and temperature, for correct data analysis. In contrast, complex
orbits such as $2e+2h$ have significant contribution for clean compounds with
moderate MB field at low temperature, only. Similar conclusions can be derived
regarding the component with frequency $F_e$ for which the contribution of $e+h$
is always significant in the range explored, except for large MB fields, while
$3e+h$ is negligible at high temperature and large scattering rate. Of course, all the considered amplitudes also depend on the field- and temperature-independent spin damping factor $R^s_{p \eta}$ through the product $g^*_{\eta}m_{\eta}$.

\section{Results and discusion}

\begin{figure} 
\centering
\resizebox{0.75\columnwidth}{!}{
  \includegraphics{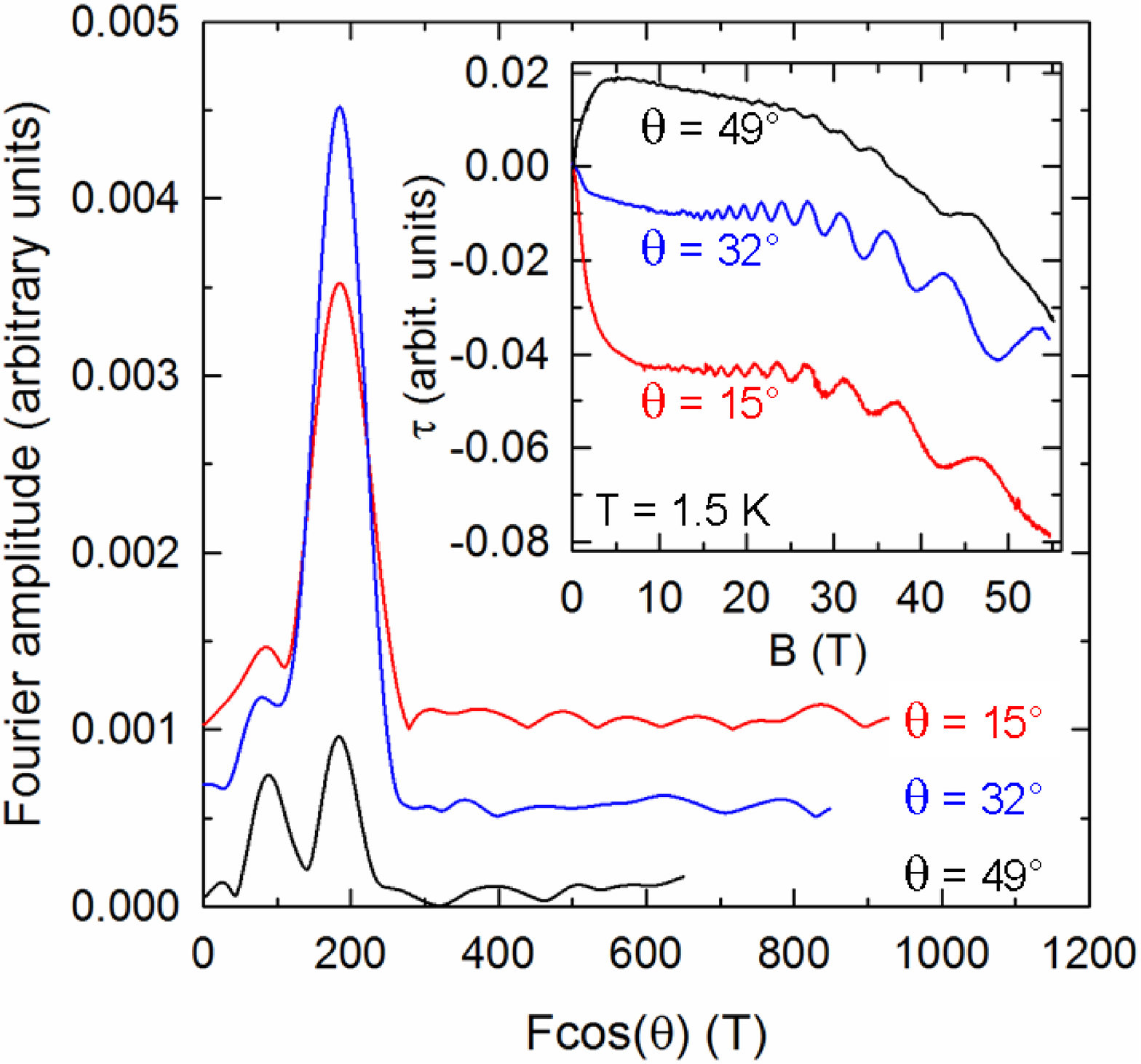}
}
\caption{\label{Fig:mm_TF} Fourier analysis in the field range 25-55 T of the
magnetic torque data reported
in the insert for various directions of the
magnetic field with respect to the normal to the conducting plane (angle $\theta$).  }
\end{figure}

\begin{figure} 
\flushright
\resizebox{0.75\columnwidth}{!}{
  \includegraphics{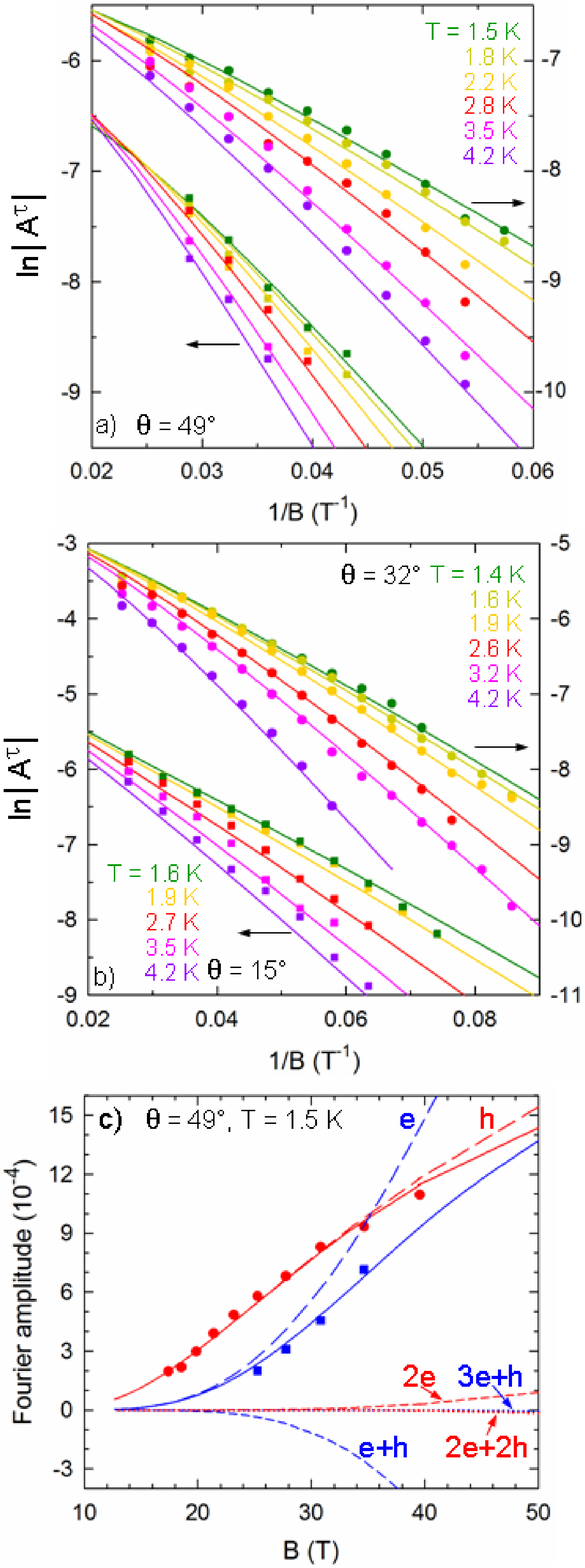}
}
\caption{\label{Fig:Dingle} ln($|A^{\tau}|$) vs 1/$B$ at various temperatures for
(a) $A^{\tau}_{h}$ (solid circles) and $A^{\tau}_{e}$ (solid squares)  at
$\theta$ = 49$^{\circ}$, (b) $A^{\tau}_{h}$ at $\theta$ = 15$^{\circ}$ (solid
squares) and 32$^{\circ}$ (solid circles) and  (c) field dependence of $A^{\tau}_{h}$ (solid circles) and $A^{\tau}_{e}$ (solid squares) at $\theta$ = 49$^{\circ}$ at 1.5 K. Solid lines are  fits of
Eqs.~\ref{eq:Atauip}, obtained with $m_e$ = 0.93,
$m_h$ = 0.88, $g^*_e$ = 2.27, $g^*_h$ = 2.06, $T_{De}$ = 5 K, $T_{Dh}$ = 4 K and $B_0$ = 2 T. Long dashed, short dashed, dotted and dash-dotted lines
in (c) are the contributions of the various components entering the fittings (see text and Fig.~\ref{Fig:calculs}).}
\end{figure}

This section is devoted to de Haas-van Alphen oscillations of
$\alpha$-'pseudo-$\kappa$'-(ET)$_4$H$_3$O[Fe(C$_2$O$_4$)$_3$]
$\cdot$(C$_6$H$_4$Br$_2$), the FS of which (see Fig.~\ref{Fig:structures}(b)) is
relevant to the above calculations. Note that this FS can be considered as resulting from the hybridization of a series of ellipses centered at the $\Gamma$ point. The area of these ellipses, which is related to the total number of holes per unit cell in the HOMO bands, is twice the area of the cross section of the FBZ because the repeat unit of the layer contains eight ET molecules with an average charge of +1/2, i.e. a total of four electrons per repeat unit. It is interesting to point out how this FS differs from that of other $\alpha$-type ET salts, such as the $\alpha$-(ET)$_2$[MHg(SCN)$_4$] family. As pointed out by Mori et al. \cite{Mo90}, the closed and open portions of the latter can also be considered to arise from the hybridization of a series of ellipses with an area equal to the cross section of the first Brillouin zone (FBZ) and centered at Y (using the axes notation of Fig.~\ref{Fig:structures}(b)). Since the repeat unit of these salts contains only four ET molecules, the FS of Fig.~\ref{Fig:structures}(b) is just a folded version along the $b^*$ direction of such FS. The different kind of overlap of the ellipses generated by the folding thus leads to the very different FS for the present salt (referred to as case (ii) above) and that of the $\alpha$-(ET)$_2$[MHg(SCN)$_4$] family (referred to as case (i) above).

Field-dependent magnetic torque data and corresponding Fourier analysis are reported in Fig.~\ref{Fig:mm_TF}. Angle dependence of the two observed frequencies (not shown) follows the cosine law predicted for a 2D FS with  $F_h(\theta=0)$  = 183 $\pm$ 3 T and $F_e(\theta=0)$ = 91 $\pm$ 5 T. Consistently with the FS topology reported in Fig.~\ref{Fig:structures}(b), they correspond to the hole and electron orbits cross section area $S_h$ = 8.9 $\pm$ 0.2 $\%$ and  $S_e$ = 4.4 $\pm$ 0.3 $\%$, respectively, of the FBZ area. These data are in agreement with magnetoresistance data ($S_h$ = 8.9 $\%$ of the FBZ area) and band structure calculations at room temperature ($S_h$ = 7.6 $\%$ of the FBZ area and $S_e$=$S_h$/2) \cite{Zo11}. Compared to the Fourier component with the frequency $F_h$, that with the frequency $F_e$, which was not observed in Ref.~\cite{Zo11}, has a relatively small amplitude at $\theta$ = 15$^{\circ}$ and 32$^{\circ}$, and can only be reliably studied at $\theta = 49^{\circ}$.

It must be noticed that the studied compound contains magnetic Fe$^{3+}$ ions. As reported in the case of $\lambda$-(BETS)$_2$FeCl$_4$ (where BETS stands for bis-ethylenedithio-tetraselenafulvalene) \cite{Uj02,Ce02}, these magnetic ions induce an exchange field leading to angle-dependent splitting of the oscillation frequency which may alter the oscillations amplitude. Nevertheless, such a splitting is not observed neither in the considered compound nor in the compound with the same composition and $\beta$'' structure in fields of up to 55 T \cite{Vi09,Vi10,La11}. This could be due to the lower, by a factor of two, Fe$^{3+}$ concentration in the present case hence to a reduced exchange coupling constant.

A naive analysis of the temperature dependence of $A^{\tau}_h$, relevant to
$F_h$, assuming that only one orbit contributes (i.e. through Eq.~\ref{eq:Mosc}) yields
an  effective mass $m_{h}(\theta) \times \cos(\theta)$ of  0.96 $\pm$ 0.05,
0.93 $\pm$ 0.02 and 0.69  $\pm$ 0.07 at $\theta$ = 15$^{\circ}$, 32$^{\circ}$
and 49$^{\circ}$, respectively. Within the same hypothesis, the magnetoresistance
data at $\theta$ = 0$^{\circ}$ of Ref.~\cite{Zo11} yield $m_{h}$ = 1.11 $\pm$ 0.04 \cite{Fn1}. In other words, the product $m_{h}(\theta) \times \cos(\theta)$ would monotonously decreases as $\theta$ increases. Hence, at variance with the angle dependence of the frequency,
the cosine law ($m_h(\theta) = m_h(0)/\cos\theta$) consistent with a
2D FS, would not be followed for the effective mass within this
assumption. This result strongly suggests that Eq.~\ref{eq:Mosc} is unable to account for the temperature dependence of the amplitude. Besides, it can be checked that the field dependence of the amplitude cannot be accounted for by Eq.~\ref{eq:Mosc} as well. As discussed in the preceding section and reported hereafter, other
orbits (such as $2e$, $2e+2h$, $etc.$) with effective masses different from each other enter the oscillatory spectra.

According to Eqs.~\ref{eq:Mosc2}, oscillatory torque amplitudes  $A^{\tau}_{e}$ and $A^{\tau}_{h}$, of the Fourier components with frequencies $F_e$ and
$F_h=2F_{e}$ observed in Fig.~\ref{Fig:mm_TF}, are given by:

\begin{eqnarray}
\label{eq:Atauip}
A^{\tau}_{e} = \tau_0 \tan(\theta) B\frac{F_e}{\pi}A_e,\;
A^{\tau}_{h} = \tau_0 \tan(\theta) B\frac{2F_e}{\pi}A_{2e},
\end{eqnarray}

respectively, where $\tau_0$ is a prefactor depending on the cantilever
stiffness, crystal geometry, $etc.$ and $\theta$ is the angle between the field
direction and the normal to the conducting plane. Amplitudes $A_{e}$  and $A_{2e}$ in Eqs.~\ref{eq:Atauip} are given by Eqs.~\ref{eq:Ae} and~\ref{eq:A2e}, respectively. We will
limit ourselves to small $n_h$ and $n_e$ values. Namely, in addition to the
basic electron and hole orbits, only the second harmonic of the electron orbit
($2e$) and the MB orbits composed of one hole and one electron orbit ($e+h$), 3
electron and one hole orbits ($3e+h$), 2 electron and 2 hole orbits ($2e+2h$) and  4 electron and one hole orbits ($4e+h$)
are taken into account. In short, the contribution of the MB orbits and harmonics composed of
more than 5 individual orbits are neglected.
It must be kept in mind that spin damping factors may influence the sign of a given contribution or, in other words, induce a $\pi$ dephasing, as observed on either side of a spin zero angle \cite{Si00}.

\begin{figure}[h]                                                    
\centering
\resizebox{0.75\columnwidth}{!}{
  \includegraphics{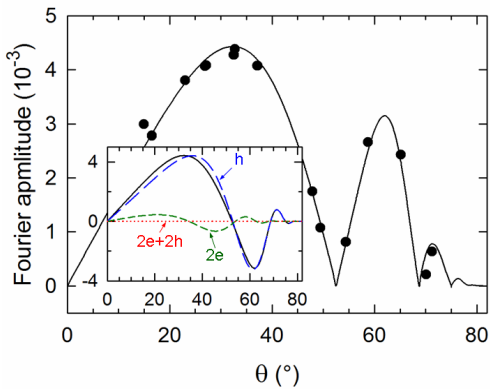}
}
\caption{\label{Fig:spin0} Angle dependence of the amplitude $|A^{\tau}_h|$ at T = 1.5 K and B = 40 T.
Solid line is the best fit of Eq.\ref{eq:Atauip} to the data, obtained
with the same set of parameters as in Fig.~\ref{Fig:Dingle}. Long dashed, short
dashed and dotted lines in the insert are the contributions of $h$, $2e$ and
$2e+2h$, respectively. }
\end{figure}

Fig.~\ref{Fig:Dingle} displays field
dependence at various temperatures of either
$A^{\tau}_{e}$ or $A^{\tau}_{h}$ for $\theta$ =  15$^{\circ}$ and 32$^{\circ}$ (crystal \#1) and 49$^{\circ}$ (crystal \#2) while Fig.~\ref{Fig:spin0} displays the angle dependence of $A^{\tau}_{h}$ at T=1.5 K and B = 40 T for crystal \#2. Solid lines in these figures are the best fits of Eqs.~\ref{eq:Atauip} to the Fourier amplitudes: the same set of parameters stands for
all the data, namely $m_e$ = 0.93 $\pm$ 0.04, $m_h$ = 0.88 $\pm$ 0.04, $g^*_e$ = 2.27  $\pm$ 0.12, $g^*_h$ = 2.06  $\pm$ 0.09. Dingle temperatures and MB field, which jointly govern the field dependence of the amplitude are obtained with a large uncertainty. Dingle temperatures are rather large, of the order of few K, whereas $B_0$ is in the range between 0 and 4 T which is rather small, in agreement with the FS of Fig.~\ref{Fig:structures}. Owing to the small  size of $e$ and $h$ orbits, compared to that of the FBZ, effective mass values are large which suggests significant renormalization due to many-body effects.

According to Dharma-wardana et al. \cite{Dh05}, electron correlations are predicted to yield large Land\'{e} factor. Consistently, large values are observed, although they are within the spread range usually reported for organic metals \cite{Wo96}. In that respect, puzzling data can be found in the literature since, for example, values as small as $g^*_{\alpha}$=1.6 and $g^*_{\beta}$=1.5 are observed for the strongly correlated $\kappa$-(ET)$_2$Cu(SCN)$_2$ compound \cite{Gv04}. Oppositely  values larger than 2 are reported for e.g. $\kappa$-(ET)$_2$I$_3$ ($g^*_{\beta}$=2.27 \cite{He93}).

Composites orbits, $e+h$ and $2e$ have significant contributions to $A^{\tau}_e$ and $A^{\tau}_h$, respectively, as observed in Fig.~\ref{Fig:Dingle}(c). Nevertheless, higher order terms (linked to $2e+2h$ and $3e+h$) are small, due to rather large
Dingle temperatures. Finally, as reported in Fig.~\ref{Fig:Dingle}c for the data at 49$^{\circ}$, contributions of $e$ and $3e+h$ to $A_e$ have opposite signs. This result which also holds at 15$^{\circ}$ and 32$^{\circ}$, explains why this amplitude is small compared to $A^{\tau}_h$.

\section{Summary and conclusion}

Fourier spectra of de Haas-van Alphen oscillations of compensated two-dimensional metals with Fermi surface composed of one hole and two electron components have been considered. The two main Fourier components observed have frequencies $F_e$ and $F_h$ = 2$F_e$, corresponding to the electron and hole orbits area. Nevertheless, it is demonstrated that, taking into account the opposite sign of electron and hole orbits, a given Fourier component involves combination of several orbits, the contribution of which must be included in the calculations. Such FS, which is a textbook case, is observed in the strongly 2D organic metal $\alpha$-'pseudo-$\kappa$'-(ET)$_4$H$_3$O[Fe(C$_2$O$_4$)$_3$]$\cdot$(C$_6$H$_4$Br$_2$). Magnetic torque oscillations of this compound have been studied up to 55 T for various directions of the magnetic field with respect to the conducting plane. It is demonstrated that data analysis performed assuming that only single electron and single hole orbits contribute to Fourier components with
frequency $F_e$ and 2$F_e$, respectively, cannot account for the data. In other words, additional orbits generated by tunneling and reflection at magnetic breakdown junctions must be taken into account. Calculations are in good quantitative agreement with the data.

\begin{acknowledgement}
This work has been supported by EuroMagNET II under the EU Contract No. 228043, and MINECO-Spain (Projects FIS2012-37549-C05-05 and CSD 2007-00041).
\end{acknowledgement}


\begin{thebibliography}{00}
\bibitem{Fn1} The, even larger, effective mass value $m_h$ = 1.26, reported in Ref.~\cite{Zo11} is deduced from a high temperature asymptotic approximation of
the temperature damping factor. 
\end{thebibliography}
\end{document}